\documentclass[aps,prl,twocolumn,superscriptaddress]{revtex4-1}
\usepackage{graphicx}
\usepackage{bm}
\usepackage{amsmath}
\usepackage{verbatim}
\usepackage{amssymb}
\usepackage{listings}
\usepackage{color}
\usepackage{epstopdf}
\renewcommand\vec[1]{\boldsymbol{#1}}

\newcommand\Fig[1]{Fig.~\ref{#1}}
\newcommand\quot[1]{``#1''}
\renewcommand\vec[1]{\boldsymbol{#1}}
\newcommand\eq[1]{Eq.~(\ref{#1})}

\let\ifincludesupplements\iftrue
\let\ifnotbuildingseparatesupp\iftrue

\begin{document}

\title{
Solid--liquid transition of skyrmions in a two-dimensional chiral magnet
}

\author{Yoshihiko Nishikawa}
\email{nishikawa@huku.c.u-tokyo.ac.jp}
\affiliation{Department of Basic Science, Graduate School of Arts and Sciences, 
The University of Tokyo, 3-8-1 Komaba, Meguro, Tokyo 153-8902, Japan}

\author{Koji Hukushima}
\email{hukusima@phys.c.u-tokyo.ac.jp}
\affiliation{Department of Basic Science, Graduate School of Arts and Sciences, 
The University of Tokyo, 3-8-1 Komaba, Meguro, Tokyo 153-8902, Japan}
\affiliation{Center for Materials Research by Information Integration,
National Institute for Materials Science, 1-2-1, Sengen, Tsukuba, 
Ibaraki, 305-0047, Japan} 

\author{Werner Krauth}
\email{werner.krauth@ens.fr}
\affiliation{Laboratoire de Physique Statistique, D\'{e}partement de physique
de l'ENS, Ecole Normale Sup\'{e}rieure, PSL Research University, Universit\'{e}
Paris Diderot, Sorbonne Paris Cit\'{e}, Sorbonne Universit\'{e}s, UPMC
Univ. Paris 06, CNRS, 75005 Paris, France}
\affiliation{
Department of Physics, Graduate School of Science,
The University of Tokyo, 7-3-1 Hongo, Bunkyo, Tokyo 113-0033, Japan
}

\date{\today}
\ifnotbuildingseparatesupp
\begin{abstract}
We study the melting of skyrmions in a two-dimensional Heisenberg chiral
magnet with bi-axial Dzyaloshinskii--Moriya interactions. These topological
excitations may form at zero temperature a triangular crystal with long-range
positional order. However, we show using large-scale Monte Carlo simulations
that at small finite temperature, the skyrmions rather form a typical
two-dimensional solid: Positional correlations decay with distance as power
laws while the orientational correlations remain finite. At higher temperature,
we observe a direct transition from this two-dimensional solid to a liquid
with short-range correlations. This differs from generic two-dimensional 
homogeneous particle systems, where a hexatic 
phase is realized between the solid and the liquid.
\end{abstract}


\maketitle
\fi

In recent years, spin textures such as magnetic skyrmions
or chiral solitons were observed in chiral magnets governed
by antisymmetric Dzyaloshinskii--Moriya (DM) interactions
\cite{Dzyaloshinsky1958,Moriya1960,Moriya1960a}.  In magnetic
systems, skyrmions were proposed theoretically as local spin vortices
\cite{Bogdanov1989,Bogdanov1994,Rossler2006} that are characterized by a
topological charge. These fascinating topological excitations have since been
extensively studied in experiments and through theoretical models.
In three-dimensional bulk compounds such as MnSi, skyrmions are confined
to a small region of the phase diagram, and they only seem to be stable
in a crystalline phase \cite{Kadowaki1982,Muhlbauer2009,Nagaosa2013}. In
two-dimensional thin films, on the other hand, skyrmions can be
stabilized in a wide range of temperatures and magnetic fields. A
triangular skyrmion crystal was observed at very low temperature in a
magnetic field \cite{Yi2009,Han2010,Yu2011,Seki2012,Nagaosa2013,Nandy2016}.
This crystal was reported stable at finite temperature
\cite{Han2010,Yu2010,Yu2011,Seki2012,Nagaosa2013,Kong2013}. Two-dimensional
skyrmions also exist as isolated objects, and they can form liquids
\cite{Yu2010,Yu2011,Nagaosa2013}, at higher temperatures than the
ordered states.

In this letter, we investigate two-dimensional skyrmions in a classical
Heisenberg spin model with DM interactions. We use massive Monte Carlo 
simulations with a dedicated parallelized algorithms implemented on GPUs 
\cite{Weigel2011,Weigel2012}. Our
analysis relies on the analogy of the skyrmion system with interacting
particle models in two dimensions.  Two-dimensional particle systems,
in the absence of a periodic substrate, cannot crystallize at finite
temperature, that is, develop long-range positional order \cite{Mermin1966},
but they can form a solid with algebraically decaying positional correlation
functions and long-range orientational order \cite{Mermin1968}. The global
orientational order $\Psi_6$ is defined as the average over all particles
of the local orientational order $\psi_j = \left\langle \exp \left( 6 i
\theta_{jk}\right) \right\rangle_j$ where the bracket represents the average
over neighboring particles of particle $j$ and $\theta_{jk}$ is the bond angle
between two particles $j$ and $k$ measured from a fixed axis.
This solid first melts  into a hexatic (with algebraic orientational
correlations and short-range positional correlations) and then, in a second 
stage, into a
liquid phase \cite{Bernard2011,Kapfer2015}.  In the liquid, all
correlation functions are short-ranged. The order of the  melting transitions
depends on the interaction potential. For steep power-law
interactions $U \left( r\right) = \epsilon \left( \sigma / r\right)^n$
with $n \gtrsim 6$ including the hard-core potential ($n \to \infty$),
the hexatic--liquid transition is of first order whereas for weaker
potentials it is continuous and can be described by the KTHNY theory
\cite{Kosterlitz1973,Halperin1978,Young1979,Nelson1979,Kosterlitz2016}.
Although skyrmions are flexible extended objects each comprising a large number
of spins and, strictly speaking, are metastable excitations with a finite
lifetime, they can, for the aim of our analysis, be pictured as stable point
particles.  Furthermore, because of its nature as a collective excitation
of a large number of spins on a lattice, the coupling of a skyrmion 
to the periodic underlying lattice is very weak.  From this viewpoint, the 
skyrmions in
the two-dimensional chiral magnet at low temperature can be interpreted as
two-dimensional particles on a fine-meshed periodic substrate.  At finite
temperatures, skyrmions feature quasi-long-range positional order  
for a floating solid that is decoupled from the underlying substrate,
whereas 
long-range positional order can be developped only if it is locked
to the substrate \cite{Halperin1978,Nelson1979}.  In our large systems with
$\sim 10^6$ Heisenberg spins and $\sim 10^4$ skyrmions, we indeed identify at 
zero
temperature a skyrmion-crystal state with locally
triangular order (minimally disturbed to accommodate the substrate potential).
This state is incommensurate with the substrate at all densities. We find 
that commensurate 
square-shaped skyrmion crystals have higher energy than triangular crystals, 
even in the very dilute limit near the transition field at zero temperature. 
They
are thus not realized. At finite temperature, the triangular crystal becomes a 
\quot{floating} solid with local triangular
symmetry.  We clearly identify the algebraic decay of positional correlations
in the presence of long-range orientational correlations. The very low degree
of coupling of this solid to the underlying spin system is also illustrated
by the fact that for a finite system the solid may vary its orientation with
respect to the lattice axes. 
We find at finite temperature that spin-lattice-commensurate
square skyrmion structures are unstable towards a locally triangular skyrmion 
solid 
with
minute inhomogeneities to accommodate the underlying lattice. As predicted by
theory \cite{Halperin1978,Nelson1979}, we find that the solid melts in a single 
step into a skyrmion liquid. The intermediate hexatic
phase found in particle systems is thus absent in the chiral magnet.

Our chiral Heisenberg spin model is defined by the Hamiltonian
\begin{eqnarray}
H \left( \left\{ \vec S_i \right\} \right) = -J \sum_{\left \langle i, j 
\right\rangle} \vec S_i \cdot \vec S_j
- \vec D_x \cdot \sum_i \vec S_i \times \vec S_{i + \hat x} \nonumber \\
- \vec D_y \cdot \sum_i \vec S_i \times \vec S_{i + \hat y}
- \vec h \cdot \sum_i \vec S_i,
\label{e:DM_definition}
\end{eqnarray}
where $\vec S_i$ is a three-component vector of fixed length $|\vec S_i|
=1$ on a site $i$ of the two-dimensional square lattice with periodic
boundary conditions.  In \eq{e:DM_definition}, the bracket $\left \langle
\cdot, \cdot\right \rangle$ represents all neighboring pairs.  The second
and third terms in the Hamiltonian constitute the DM interaction.  We take
$\vec h$ parallel to the $z$-axis, $\vec D_x$ parallel to the $x$-axis, $\vec
D_y$ parallel to the $y$-axis, and choose $| \vec D_x | = | \vec D_y | = J$.  
This choice
stabilizes Bloch-type skyrmions at low temperature in a magnetic field
\cite{Kezsmarki2015}.  This model has helical, paramagnetic (skyrmion-liquid)
and skyrmion-solid
phases depending on temperature and the magnetic field.  We especially focus on
the skyrmion solid at low temperature and the higher magnetic
field (see \Fig{fig:fig1}).

\begin{figure}[bp]
\includegraphics[width=0.75\linewidth]{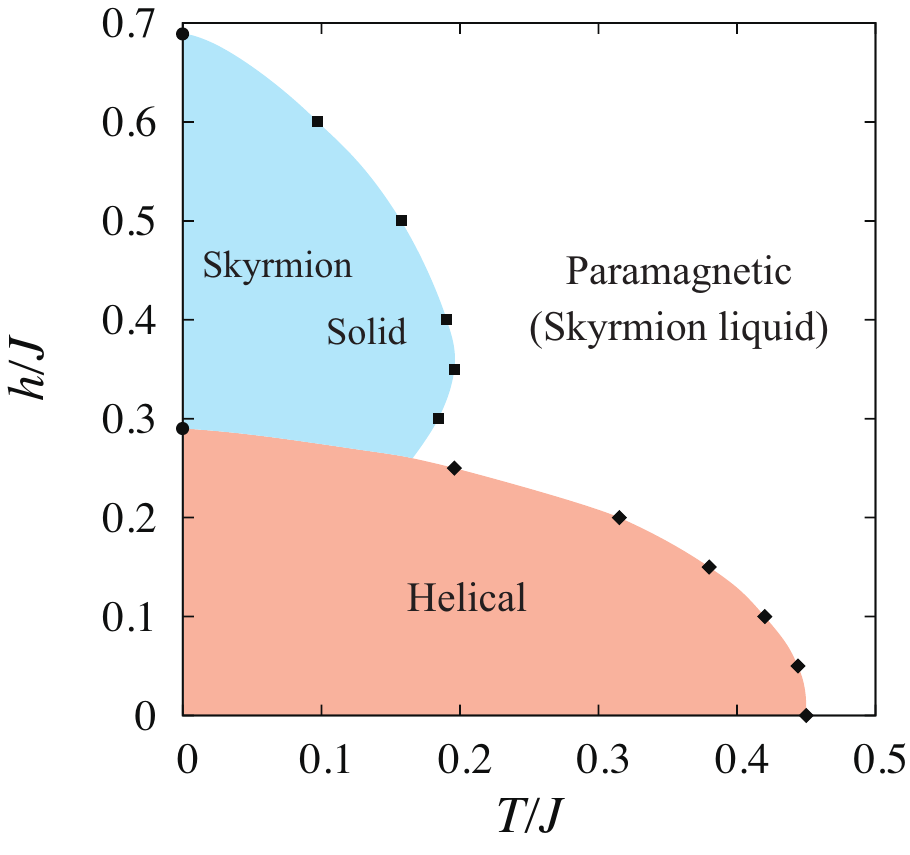}
\caption{\small
Magnetic phase diagram of the two-dimensional chiral magnet.
Circles correspond to zero-temperature energy minimizations.
Squares are obtained by decays of correlation 
functions, and diamonds denote peak locations
of the specific heat obtained by finite-temperature Monte Carlo simulations
at each magnetic field.
}
\label{fig:fig1}
\end{figure}

With the parameter $D / J = 1$, each skyrmion is composed of $\sim 50$ spins, 
and encounters large free-energy barriers with respect to local changes as they 
appear
in experiment and in  local sampling algorithms. On the one hand, this 
large 
free-energy barrier stabilizes skyrmions at low temperatures
on experimental lifetimes. On the other hand, this complicates Monte Carlo 
calculations as at low temperature, in the solid phase, the skyrmion number 
$N_\mathrm s$, although not strictly stable, 
changes too slowly to allow its efficient sampling by our algorithms. 

To overcome this problem, our Monte Carlo simulations are performed in two 
stages. 
In the first stage, we determine the thermodynamic relevant value of skyrmions 
$N_\mathrm s$ (equivalent to the skyrmion density) at a low target temperature 
$T$ 
through
extensive simulated annealing runs from high temperature (where $N_\mathrm s$ 
changes easily) down to $T $.
In the second stage, we compute correlation functions 
using Monte Carlo simulations where, in contrast, we keep the skyrmion number 
rigorously fixed to the thermodynamic relevant value. To do so, we 
compute
$N_\mathrm s$ after every Monte Carlo time step (using the topological charge 
\cite{Berg1981}), and reject configurations with changed 
skyrmion 
number (see the Supplemental Item 1 for details on the 
Monte 
Carlo procedure at zero and at finite temperatures).
Using a local mask, we 
identify the center of each skyrmion as a point particle with real-valued 
positions and then compute 
high-quality spatial correlation functions at fixed $N_\mathrm s$ (see 
Supplemental Item 3 for a 
description of our algorithm to determine real-valued skyrmion positions). 
For the latter analysis, it is essential that different values of  $N_\mathrm 
s$ are not folded into the analysis of correlation functions. 

\begin{figure}[b]
\includegraphics[width=\linewidth]{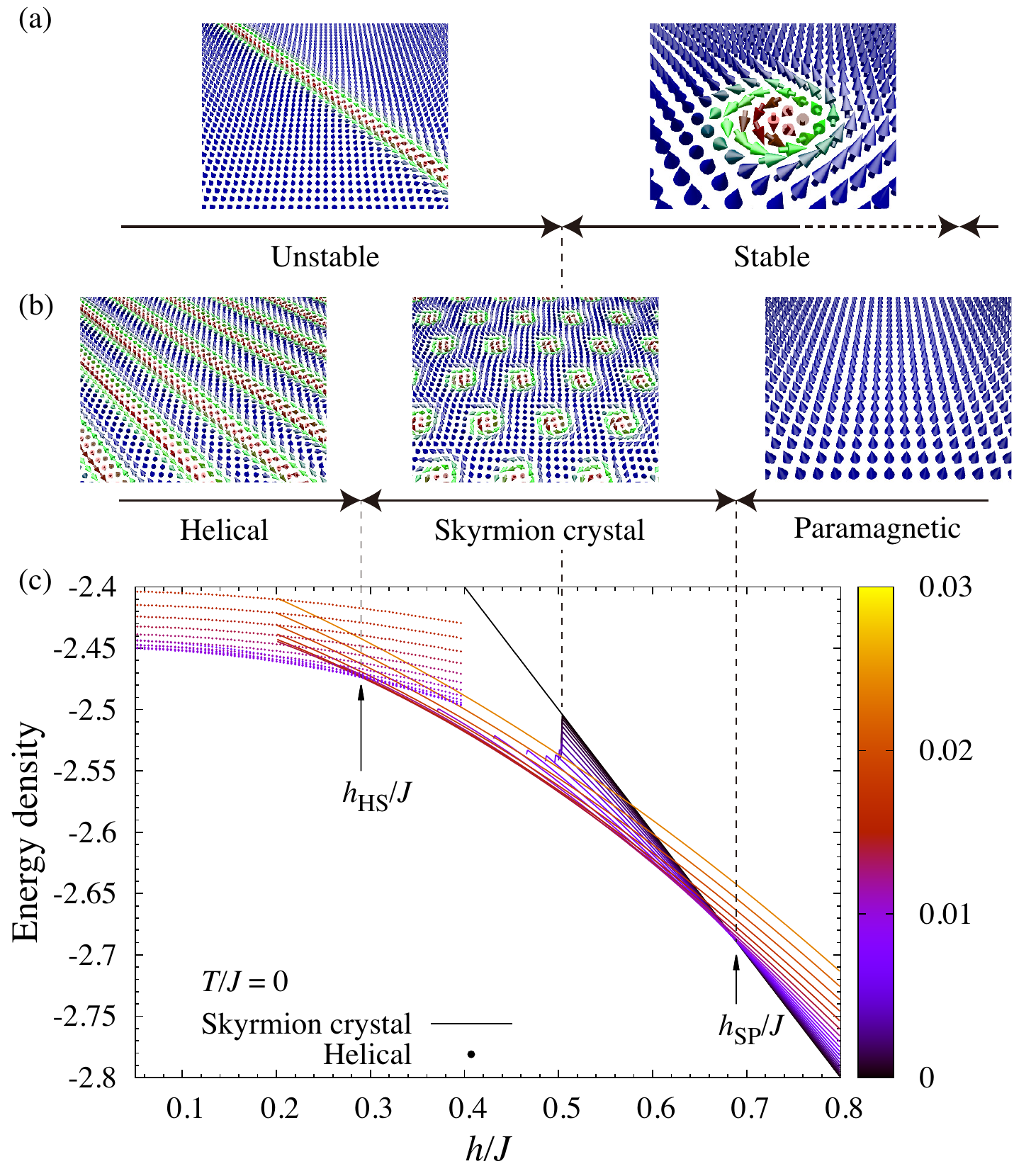}
\caption{(a) Phase diagram of a single skyrmion at zero temperature.
For $h / J \lesssim 0.50$, a 
single skyrmion is unstable towards a tube-like object, and at high $h/J 
\gtrsim 1.10$ towards the paramagnetic state. (b) Phase diagram of the system 
at zero temperature, with helical, triangular skyrmion crystal, and 
paramagnetic state ($h_\mathrm{HS} / J \simeq 0.29$ and $h_\mathrm{SP} / J 
\simeq 0.69$).
(c) Energy densities of triangular skyrmion crystal states (with various 
densities) and helical states (with various winding numbers). The color 
represents the skyrmion densities for skyrmion crystal states, and square of 
winding densities for helical states, respectively. Square skyrmion crystals 
always have higher energies than the triangular skyrmion-crystal state of 
lowest energy.}
\label{fig:fig2}
\end{figure}

\begin{figure*}[t]
\includegraphics[width=\linewidth]{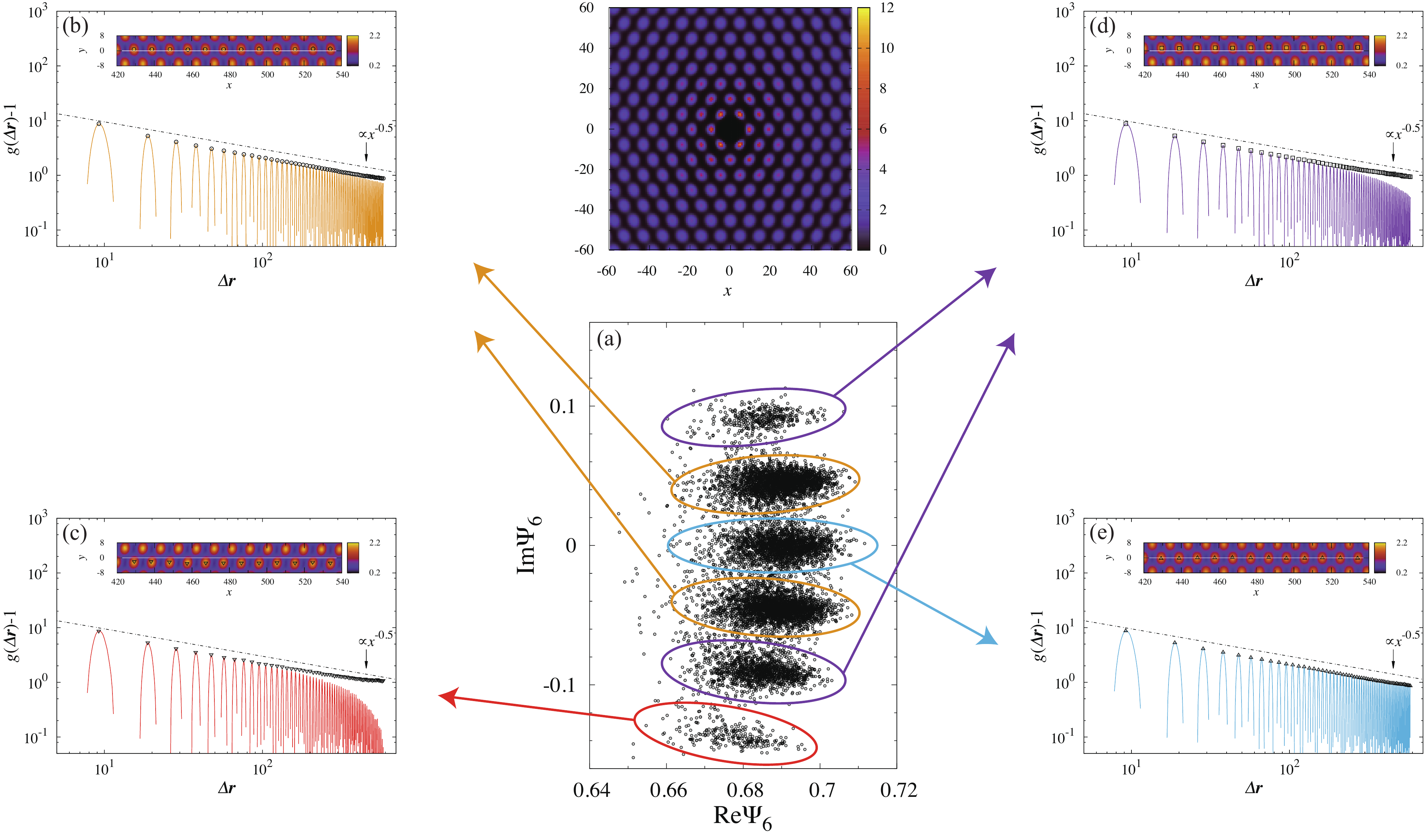}
\caption{(a) Scatter plot of the global bond-orientational order parameter
$\Psi_6$, and the two-dimensional pair correlation function. (b)--(e) 
Positional correlation function $g \left( \Delta \vec r \right) - 1$ for each 
lobe of the scatter plot. Solid lines represent $g\left( x, 0 \right) - 1$ 
and black open symbols (circles in (b), inverted triangles in (c), etc.) 
represent the maximum value of each peak in the two-dimensional 
correlation function $g \left( x, y \right)$. Insets in (b)--(e) show the 
two-dimensional correlation function $g \left( x, y \right)$ near the $x$ axis 
(compare with (a)). The white line is the $x$ axis along which $g\left( x, 0 
\right) - 1$ is plotted. Black open symbols again represent local maxima. The 
system is $L_x = 1164$ and $L_y = 1008$ at $T / J = 0.155$ and $h / J = 0.5$.
}
\label{fig:fig3}
\end{figure*}

\begin{figure}[t]
\includegraphics[width=\linewidth]{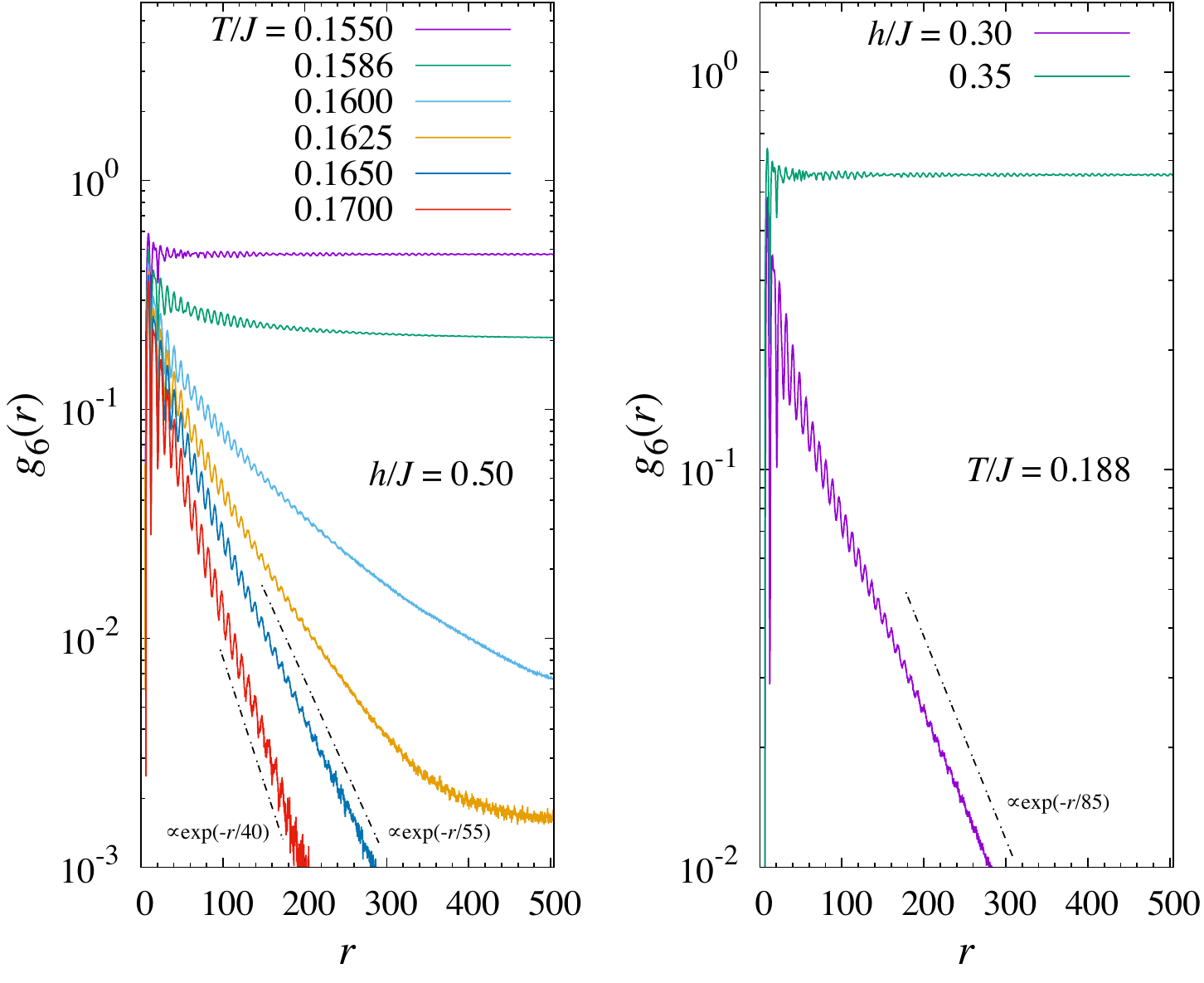}
\caption{Correlation function of the orientational order $g_6 (r)$ 
of the system with $L_x = 1164$ and $L_y = 1008$ at various temperatures.
The magnetic field is (a) $h / J = 0.5$, and (b) $h / J = 0.30$ and $0.35$.
}
\label{fig:fig4}
\end{figure}

Solid and crystalline skyrmion states are sensitive to the boundary conditions, 
that is, to the shape of the simulation box. We therefore carefully choose the 
linear 
dimensions of the spin lattice in order to minimize the distortion of solid and 
crystalline states (see the Supplemental Item 1 for details of our Monte Carlo 
procedure and of the choice of system size).

At zero temperature, a single skyrmion is stable above $h / J \simeq
0.50$ and below $h/J \simeq 1.10$. However, the paramagnetic 
solution \footnote[1]{The paramagnetic phase at zero temperature is often incorrectly 
referred as the ferromagnetic phase} (which, at $T / J = 0$, is without skymions) has lower 
energy than the single-skyrmion solution above $h_\mathrm{SP} / J = 0.69$.
Below $h / J \simeq
0.50$, a single skyrmion is unstable and transforms into another tube-like 
object with the
same topological charge (see \Fig{fig:fig2}(a)). 
At $h / J \lesssim 0.50$, very low-density skyrmion states
also become unstable, but they have higher energy than
skyrmion crystal states of higher density.
The latter are in fact stabilized by the mutual repulsion
of skyrmions (see also 
\cite{Yi2009,Han2010,Yu2011,Nagaosa2013,Iwasaki2013,Nandy2016}). 
Above $h_\mathrm{HS} / J \simeq 0.29$, skyrmion crystal states thus have lower energy 
than the helical states of various periods (see \Fig{fig:fig2}(c)).
We also find that square-lattice crystal configurations have higher energy than 
triangular lattices that minimally adjust to the substrate potential. 
All these crystalline and helical states are obtained by simulated 
annealing (SA) in magnetic field
(see Supplemental Item 1 for details on the simulated annealing
at zero temperature). We thus
conclude at zero temperature that a triangular skyrmion crystal state (with
minimal local variations) is the ground state between $h_\mathrm{HS} / J$ and
$h_\mathrm{SP} / J$, see \Fig{fig:fig2}(b). 

At finite temperature, the correct skyrmion density at low temperature is 
determined by simulated annealing from high temperature (see Supplemental
Item 2 for details on this simulated annealing procedure). For 
$T / J = 0.155$ and $h / J = 0.5$, for example, we estimate the correct number 
to be $128 \times 128$ for $L_x = 1164$ and $L_y = 1008$. This value remarkably 
corresponds to the zero-temperature value. In order to determine 
spatial correlations of orientational and positional order, as described, we 
thus freeze the 
skyrmion number to the value found by simulated annealing. 
At low temperature, the global orientational order $\Psi_6$ may be in 
different 
\quot{lobes} that correspond to slight rearrangements of the skyrmion system 
with respect to the 
simulation box (see \Fig{fig:fig3}(a)). It is only at near-zero temperature 
that the number of sampled lobes in the histogram shrinks to one, as
the orientational order is then locked into the value of the initial 
configuration. We compute the two-dimensional positional correlation 
function $g \left( x, y \right)$ separately for each lobe, as each of the 
orientations leads to distinct distortions. (Each configuration is rotated by 
$-\arg \left( \Psi_6 \right) / 6$ so that the resultant $\Psi_6$ is 
approximately parallel to the $x$ axis \cite{Bernard2011}). The positional 
correlation function $g \left(x, 0 \right) - 1$ for the lobe at $\mathrm{Im} 
\Psi_6 \simeq 0$, at $T/J = 0.155$ and $h/J = 0.5$ clearly decays 
algebraically, with an exponent about $0.5$ 
(\Fig{fig:fig3}(e)). At first sight, the 
positional correlation functions 
for other lobes  appear to decay 
faster than algebraic at large distance.  However, this is an 
artifact due to the slight distortion of the correlation peaks near the $x$
axis that is caused by the mismatch of the locally triangular structure  
with the simulation box if $\mathrm{Im} \Psi_6 \neq 0$ (see the insets of 
\Fig{fig:fig3}(b)--(e)).  
The maximum values for all peaks, determined individually, indeed
decay algebraically for all the lobes (see black open symbols in 
\Fig{fig:fig3}(b)--(d)).
The correlation function of the orientational order parameter
$g_6 \left( r \right)$ at this temperature converges to a constant value at a
large distance that means $g_6 \left( r \right)$ has long-range correlations
(\Fig{fig:fig4}(a)).  These orientational correlations are also observed in a 
smaller system
with $582 \times 504$ spins and $64 \times 64$ skyrmions, and they 
are consistent with
an algebraic decay with exponent $\simeq 0.5$ in the positional correlation
function. Therefore we conclude that skyrmions do not form a crystal but a
solid at low temperature.  While the positional correlation function shows a
clear algebraic decay, the exponent $0.5$ is larger than the exponent $1/3$
that is the stability limit of the solid phase predicted by the KTHNY theory.
Our system has no continuous symmetry, and each position of skyrmions has small
but finite coupling to the lattice sites (see Supplemental Item 4 for details
on the coupling between the skyrion positions and the substrate). This coupling
can be considered as an effective periodic potential incommensurate to the
skyrmion solid that stabilizes the solid phase with a larger 
exponent
than that observed in particle systems. A larger exponent was also reported
in an experimental work on a melting transition of atoms on a two-dimensional
periodic substrate \cite{Negulyaev2009}.

In the solid phase, where the positional correlations decay algebraically 
(see \Fig{fig:fig3}), orientational correlations $g_6 \left(r \right)$ are 
long-ranged, whereas, in the liquid, orientational correlations (as all other 
correlations) decay exponentially. Near the transition between the two phases, 
we observe, from the solid, a decrease of the asymptotic correlation $g_6$, and 
from the liquid a very rapid increase of the orientational correlation length 
towards a value on order of the system size $L$
(see, for example, \Fig{fig:fig4}(a) at $h/J = 0.5$, where the transition 
is at $T/ J \simeq 0.16$). Analogous behavior is found for all other 
values of $h/J$ with a solid--paramagnet transition. Unlike for 
the two-dimensional particle models, but in agreement with theoretical 
predictions
\cite{Halperin1978,Nelson1979}, we find no evidence of a hexatic phase.
This is also consistent with the experimental work
\cite{Negulyaev2009} that reports a direct melting transition from the solid
to the liquid. Typically, 
the skyrmion solid melts into the paramagnet with increasing $h/J$, as
the paramagnetic state favors magnetic order. However, 
in a small region of the phase diagram, the paramagnet is 
switched from the paramagnet into the skyrmion-solid phase on 
increasing the magnetic field (see \Fig{fig:fig1}). Accordingly, 
at a finite temperature, the orientational correlation functions are 
long-ranged at higher $h/ J$ and exponentially decaying at lower 
$h / J$ (see \Fig{fig:fig4}(b)).

In conclusion, we have numerically studied phase transitions of skyrmions in a
two-dimensional chiral Heisenberg system using massive Monte Carlo simulations
at zero and at finite temperature with a dedicated algorithm that avoids the
difficulty of relaxation to equilibrium due to the long skyrmion life time.
We confirm that the ground state of the system
is a triangular skyrmion crystal state with long-range positional order.
A spin-lattice compatible skyrmion crystal is energetically unfavorable.
However, the ordered skyrmion state at finite temperature is a \quot{floating}
solid, that is locally triangular, and has long-range orientational
correlations yet only quasi-long-range positional correlations.  The
skyrmion ground state is continuously reached from the solid as the correlation
length becomes ever larger.  On increasing the temperature, the skyrmion solid
directly melts into the liquid without an intermediate hexatic phase
as it is realized in generic two-dimensional particle systems. 
We note that the DM interaction $D / J = 1$ in our system is larger and the size of 
each skyrmion smaller than
that observed in experimental compounds \cite{Kezsmarki2015,Nagaosa2013}.
The coupling between the location of each skyrmion and the lattice sites
becomes weaker for smaller $D / J$ and larger skyrmion, and it vanishes in
the continuum model with $D / J \to 0$ limit where the melting scenario in
two-dimensional particles is expected to hold. Therefore we can expect that our
results for the low-temperature solid phase with $D / J = 1$ 
also hold for the case with smaller but finite $D / J$.

\acknowledgments{
The authors thank Yusuke Masaki for many useful discussions. 
This work was supported by a Grant-in-Aid for JSPS Fellows (Grant
No. 17J10496) and a Grant-in-Aid for Scientific Research (No. 25610102). 
}

\bibliography{skyrmion}
\ifincludesupplements

\clearpage

\title{Supplemental Material for ``Solid--liquid transition of skyrmions in a
two-dimensional chiral magnet''}

\maketitle

\onecolumngrid
\renewcommand{\theequation}{S\arabic{equation}}
\renewcommand{\thefigure}{S\arabic{figure}}
\setcounter{equation}{0}
\setcounter{figure}{0}
\setcounter{page}{1}
\setcounter{section}{0}

\subsection{Supplemental Item 1: Detailed parameters for the Monte Carlo code}

\subsubsection{Choice of lattice parameters}
For probing solid skyrmion states with locally triangular structure 
and their melting into a liquid, we choose a periodic rectangular spin lattice 
with $L_x = k \times 97$ and $L_y = k \times 84$ with integer $k$.
This realizes a ratio $L_y/L_x = 0.865979...$ very close 
to $\sqrt{3}/2 = .866025...$, and the distortion is minimal for  a triangular 
skyrmion crystal 
(in ``base'' configuration)
with $N_\mathrm{sk} = n \times n$ skyrmions with even $n$ ($n$ rows in the $y$ 
direction of $n$ skyrmions in the $x$ direction). To probe helical skyrmion 
states with $\pm 45^\circ$ tilting as well as crystal skyrmion states with 
square-lattice structure, we rather use square-shaped spin lattices with linear 
length $L$.

\subsubsection{Energy minimization and simulated annealing in magnetic field  at $T / J 
=0$}
At zero temperature, energy minimization and simulated annealing
in magnetic field 
are performed by the zero-temperature heat-bath
algorithm (where spins are aligned with their molecular 
field produced by neighboring spins and by the external magnetic field $h$).
This produces the zero-temperature phase diagram (see
\Fig{fig:fig2} of the main text).  
We typically use $4 \times 10^3$ sweeps to minimize the energy of the system at 
each magnetic field.
Triangular-crystal and square-shaped skyrmion initial configurations 
at various densities are pieced together from individual skyrmions placed
very close to the corresponding lattice sites (each individual skyrmion is 
prepared in
a small system with $7 \times 7$ spins at $h / J = 0.6$).  The helical 
initial configurations for \Fig{fig:fig2}(b)
are prepared at $h / J = 0$ from a single wave
vector $2m\pi \left(1, 1 \right) / L $ with integer $m$  and then 
relaxed 
through the zero-temperature heat-bath algorithm.  
Simulated annealing in $h$ with steps of size $|\Delta h / J|  =
0.001$ then leads from $h / J = 0.6$ to lower $h$, and from 
$h / J = 0$ to higher $h$ for
for triangular
skyrmion crystal states, and from $h / J = 0$ to higher magnetic field for
helical states (see \Fig{fig:fig2} of the main text).
This yields the energies for helical, triangular-crystal and square-crystal 
states, as well as for the paramagnetic state.
System sizes are $\left( L_x, L_y \right) = (291, 252)$
for triangular skyrmion crystals, and $L = 256$ for helical states and for 
square skyrmion crystals.

\subsubsection{Monte Carlo parameters at finite temperature}
In our computations of the correlation functions at 
finite temperature,
one unit of Monte Carlo time consists of one heat-bath sweep and of 
$10^4$ 
over-relaxation sweeps, where one sweep denotes one update per spin. We use 
GPUs to simulate the system with the 
checkerboard decomposition \cite{Weigel2011,Weigel2012}. 
Each block of the checkerboard consists of $16 \times 16$ spins. As the 
lattice dimensions $L_x$ and $L_y$ are not necessarily multiples of the 
block size, the
checkerboard is randomly shifted after every $10^2$ over-relaxation sweeps
in order to assure the ergodicity of the algorithm. 

Because of the anisotropic shape of the underlying rectangular lattice, the
number of skyrmions is incommensurate with the ``tip'' crystalline configuration
with $\mathrm{Re}\Psi_6 < 0$, and the lattice structure of $g \left(x, y
\right)$ would be strongly distorted, resulting in a higher energy. We thus 
focus on ``base'' configurations which minimize the distortion with the 
underlying spin lattice.

\subsection{Supplemental Item 2: Cooling-rate dependence of the number of 
skyrmions}
\label{supp3}

\begin{figure}[htbp]
\includegraphics[width=0.4\linewidth]{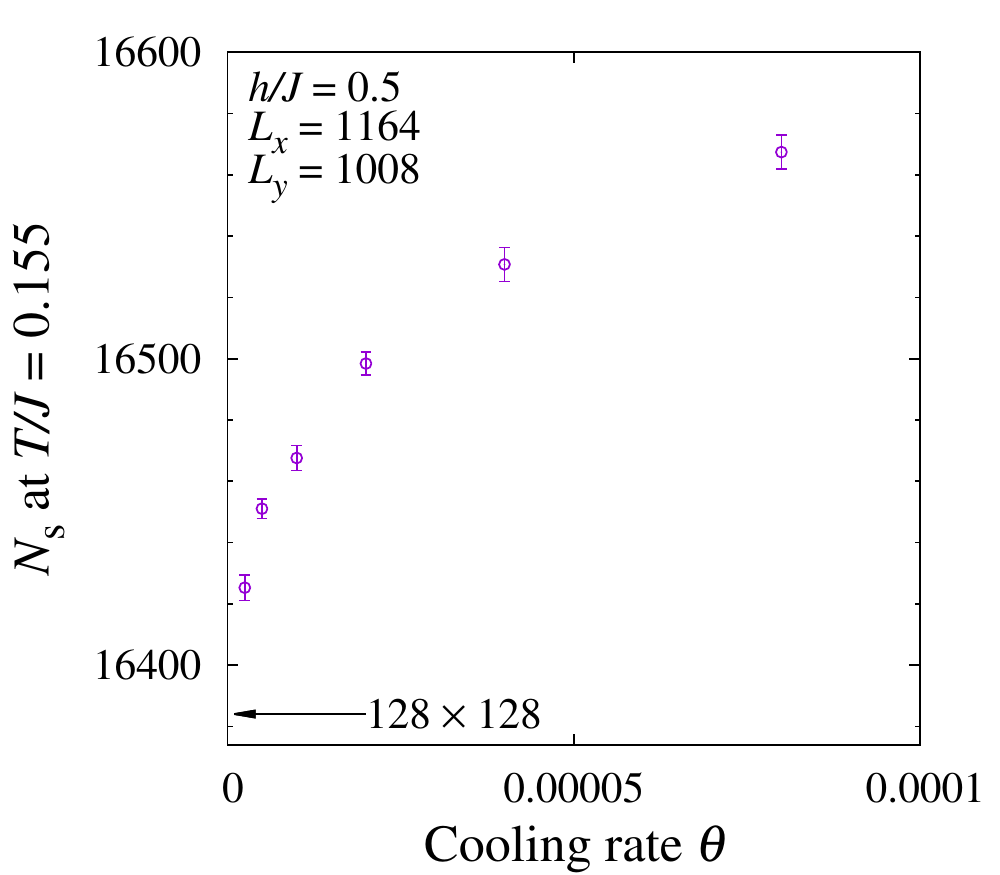}
\caption{\small
Cooling-rate dependence of the number of skyrmions in
configurations at $T / J = 0.155$ obtained by simulated annealing simulations
from a high temperature. The magnetic field $h / J = 0.5$. 
The leftmost point $\theta = 0$ corresponds to
the equilibrium limit.
}
\label{fig:figS1}
\end{figure}

In our production runs, the skyrmion number is kept rigorously fixed at what 
we believe to be the thermodynamically relevant value. 
To determine this dominant density of skyrmions at finite temperature, we run 
simulated annealing (SA) simulations from high temperature. In our SA at $h / J = 0.5$
starting from $T / J = 0.955$ to $0.155$ with $\Delta T / J = 0.01$, the number of
Monte Carlo steps at each temperature $M$ is controlled, where one Monte
Carlo time step consists of one heat-bath sweep followed by ten over-relaxation
sweeps. The resultant skyrmion number
$N_\mathrm s$ decreases with decreasing 
the cooling rate $\theta = (\Delta T / J) / M$ (see 
\Fig{fig:figS1}). The equilibrium density of
skyrmions is obtained in the limit of vanishing cooling rate. $N_\mathrm
s$ approaches $128 \times 128$ in this limit for the system with $L_x =
1164$ and $L_y = 1008$.  For smaller systems with $(L_x, L_y) = (582, 504)$ and
$(291, 252)$, the same limiting density is obtained. 

\subsection{Supplemental Item 3: Determining the position of a single skyrmion}

Although each skyrmion is composed of Heisenberg spins on a 
discrete square lattice, 
we may assign it a real-valued position ($(x,y)$).
This allows us to effectively map the Heisenberg-spin model to 
a model of interacting particles   
and to compute positional and the bond-orientational order, in analogy to what 
is done for two-dimensional particle systems.  At zero temperature, an isolated 
skyrmion has a symmetric structure around its core, and spins near a core of a 
skyrmion are antiparallel to the magnetic
field (they thus point into the $-z$ direction)). We thus consider a connected 
cluster of spins with $S_i^{\left( z
\right)} < 0$ as a skyrmion, and define the position of the skyrmion 
$\vec R_\mathrm{sk}$ in the two-dimensional plane as
\begin{equation}
R_\mathrm{sk}^{\left( \alpha \right)} = 
\frac{1}{A_\mathrm{sk}}\sum_{i \in \mathrm{skyrmion}}\left( r_i^{\left( \alpha 
\right)} 
+ S_i^{\left( \alpha \right)} \right), \quad (\alpha = x, y)
\label{e:skyrmion_locator}
\end{equation}
where $A_\mathrm{sk}$ is the number of spins composing the skyrmion. 
At finite temperature, the thermal fluctuations of the Heisenberg spins 
induce fluctuations in the determination of the skyrmion position.

\subsection{Supplemental Item 4: Coupling between skyrmion positions and 
the spin lattice}
The skyrmion locator of \eq{e:skyrmion_locator} allows us to compute the 
coupling potential between skyrmion locations and spin lattice sites. We 
simulate one
single skyrmion in the system with $16 \times 16$ spins at $T / J = 0.1$ to
obtain a histogram of the skyrmion locations
\begin{equation*}
P \left( \Delta \vec r \right) 
= \left\langle 
\delta \left( \Delta \vec r - \min_k \left(\vec R_\mathrm{sk} - \vec \ell_k 
\right)\right) \right\rangle,
\end{equation*}
where the bracket $\left\langle \cdots \right\rangle$ represents average
over configurations with only one skyrmion, and $\vec \ell_k$ represents the
location of $k$-th lattice site.  Then, the coupling potential is estimated as
$V_\mathrm{coup} \left( \Delta\vec r\right) = -\log \left( P\left( \Delta \vec
r \right) \right) / \beta$.  \Fig{fig:sk-c-l} shows $V_\mathrm{coup} \left(
\Delta\vec r\right)$ at various magnetic fields.  The coupling
potential is clearly nonzero. Furthermore, the potential minimum depends 
on the magnetic field (see
\Fig{fig:sk-c-l}). We checked that the coupling potential is independent of
system size.

\begin{figure}[b]
\includegraphics[width=1\linewidth]{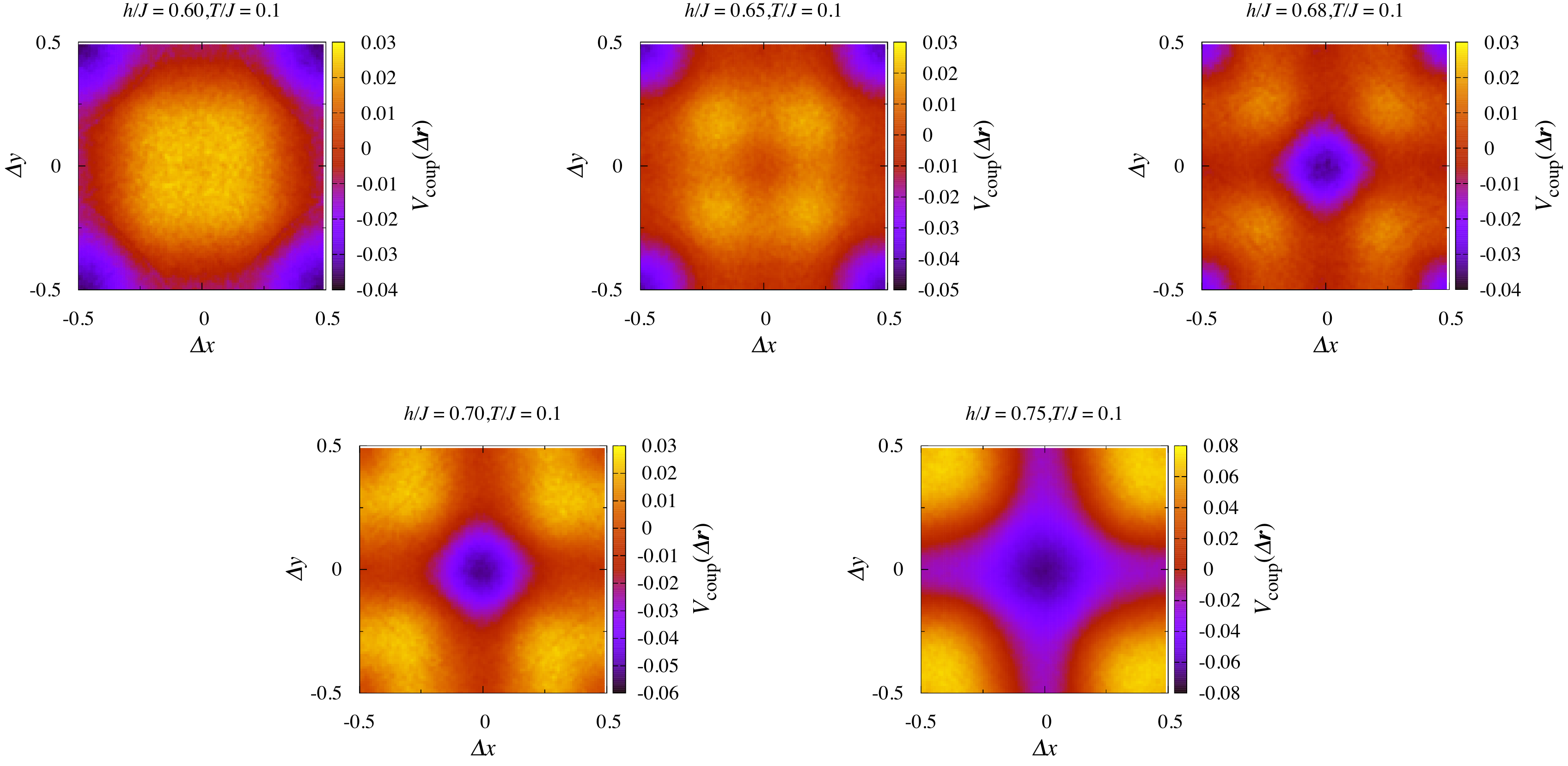}
\caption{
Effective coupling potential $V_\mathrm{coup} \left( \Delta \vec r \right)$
characterizing the coupling between skyrmion positions  and lattice
sites. For $h / J \gtrsim 0.68$, 
the potential minimum 
coincides with the  lattice sites ($\Delta x = \Delta y = 0$),
but at smaller magnetic field, the minimum lies 
between the lattice spins, at $\Delta x = \Delta y = 0.5$.
}
\label{fig:sk-c-l}
\end{figure}

\subsection{Supplemental Item 5: Phase diagram of the system}
Transition temperatures between the paramagnetic and the helical phases in
the phase diagram \Fig{fig:fig1} are estimated as the peak locations of the
specific heat. We perform regular Monte Carlo simulations using the heat-bath,
the over-relaxation, and the exchange Monte Carlo (parallel tempering)
algorithms. System sizes range from $L = 32$ (the total number of spins is
$N = 1024$) to $L = 256$ ($N = 65536$). The number of Monte Carlo sweeps is
typically $10^5$. We checked that the peak location of the specific heat varies 
very little with the system size, up to $L = 256$.

\fi
\end{document}